\documentstyle[11pt,epsf]{article}
\begin{document}
\title{A full one-loop charge symmetry breaking effective potential}
\author{P.M. Ferreira \\ Dublin Institute for Advanced Studies, \\
Ireland}
\date{November, 2000} 
\maketitle
\noindent
{\bf Abstract.} We calculate the one-loop contributions to the effective 
potential for the minimal supersymmetric model when scalar fields other than
the Higgses have vacuum expectation values. The importance of these 
contributions for studies of charge and colour breaking bounds is discussed.
\vspace{1cm}

In supersymmetric models, in order to give mass both to the up and down-type 
quarks (and also to cancel chiral and $SU(2)$ anomalies), two Higgs doublets, 
$H_1$ and $H_2$, are required. The neutral components of these fields have a 
non-zero vev which, much in the same manner as in the SM, breaks the electroweak
gauge symmetry. However, there are now other scalar fields in the game, several 
of them carrying charge, colour or leptonic number, and {\em a priori} there is 
no reason why one or many of them cannot have a vev. This problem was first 
addressed in ref.~\cite{fre} and used to impose restrictions on the undesirably 
large parameter space of supersymmetric models, by demanding that the minimum of
the potential along charge and/or colour breaking (CCB) 
directions (meaning, when fields other than $H_1^0$ and $H_2^0$ acquire vevs) be
less deep than the so-called ``real" minimum. This very appealing idea was 
developed by many authors~\cite{ccb,gu}, and of particular importance was the 
contribution from Gamberini {\em et al}~\cite{gam}: they showed that the 
one-loop corrections to the effective potential in supersymmetric theories were 
of immense relevance for gauge symmetry breaking, in that only at a given 
renormalisation scale $M$ - namely, $M$ of the order of the most significant 
mass appearing in the one-loop corrections - can their contribution to the 
minimisation conditions be neglected and results derived from the tree-level
potential trusted. Given that every work on CCB is based on an analysis of the 
tree-level potential only, they concluded that any bound on supersymmetric 
parameters we obtain in this way, if not evaluated at the scale $M$, will lead 
to an excessive constriction on the parameter space~\cite{gam}. A very 
comprehensive review of the subject is given in ref.~\cite{cas}. Recent use
of CCB bounds in supersymmetric phenomenology may be found in 
refs.~\cite{rec,euro}. In this article we will investigate the importance of
the one-loop contributions to the effective potential in determining CCB bounds,
as well as discuss the choice of renormalisation scale at which such  bounds
should be considered.

We will follow the analysis of reference~\cite{bbo}, and use most of the 
conventions of that article (as in the factors of $\sqrt{2}$ for the vevs
considered). Our convention for the sign of $\mu$ is, however, the opposite.   
The superpotential for the model we are considering (no Yukawa couplings other 
than those of the third generation) is given by
\begin{eqnarray}
W & = & \lambda_t H_2\,Q\,t_R \; + \; \lambda_b H_1\,Q\,b_R \; +\; \lambda_\tau
H_1\,L\,\tau_R \; +\; \mu H_2\,H_1 \nonumber \\
 & = & \lambda_t (H_2^+\,b_L \;-\;H_2^0\,t_L)\,t_R \;+\;\lambda_b (H_1^0\,b_L
\;-\;H_1^-\,t_L)\,b_R\; +\; \lambda_\tau(H_1^0\,\tau_L \;-\;H_1^-\,\nu_L)\,
\tau_R \; + \nonumber \\
 & & \mu \,(H_2^+\,H_1^- \;-\;H_2^0\,H_1^0) \;\;\; ,
\end{eqnarray}
The tree-level effective potential is the sum of the F, D and soft terms, which 
are given by:
\begin{eqnarray}
V_F & =& \sum_i\;\left| \frac{\partial W}{\partial \Phi_i}\right|^2 \nonumber \\
 & = & |\lambda_t\,b_L\,t_R \; + \; \mu\,H_1^-|^2 \;+\; |\mu\,H_1^0 \;+\;
\lambda_t\,t_L\,t_R|^2 \; + \nonumber \\
 & & |\lambda_b\,b_L\,b_R\;+\:\lambda_\tau\,\tau_L\,\tau_R\;-\;\mu\,H_2^0|^2 \;
+\;|\lambda_b\,t_L\,b_R\;+\;\lambda_\tau\,\nu_L\tau_R\;-\;\mu\,H_2^+|^2 \;+
\nonumber \\
 & & |\lambda_t\,H_2^0\,t_R\;+\;\lambda_b\,H_1^-\,b_R|^2\;+\;|\lambda_t\,(H_2^+
\,b_L\,\;-\;H_2^0\,t_L)|^2\;+\;|\lambda_t\,H_2^+\,t_R\;+\;\lambda_b\,H_1^0\,b_R
|^2 \;+ \nonumber \\
 & & |\lambda_b\,(H_1^0\,b_L\;-\;H_1^-\,t_L)|^2\;+\;|\lambda_\tau\,(H_1^0\,
\tau_L\;-\;H_1^-\,\nu_L)|^2\;+\;|\lambda_\tau\,H_1^0\,\tau_R|^2\;+ \nonumber \\
 & & |\lambda_\tau\,H_1^-\,\tau_L|^2 \;\;\; , \\
V_D & = & \frac{{g^\prime}^2}{2}\left[ \frac{1}{6}\,(|t_L|^2\;+\;|b_L|^2) \;-\;
\frac{2}{3}\,|t_R|^2\;+\;\frac{1}{3}\,|b_R|^2\;-\;\frac{1}{2}\,(|\tau_L|^2\;+\;
|\nu_L|^2)\;+ \right. \nonumber \\
 & & \left. |\tau_R|^2\;+\;\frac{1}{2} (|H_2^0|^2\;+\;|H_2^+|^2)\;-\;\frac{1}{2}
\,(|H_1^0|^2\;+ \; |H_1^-|^2) \right]^2 \; + \nonumber \\
 & & \frac{g_2^2}{8} \,\left[ 4\,|{H_1^*}^i\,H_2^i|^2\;-\;2\,|H_1|^2\,|H_2|^2
\;+\; 4\,|{H_1^*}^i\,Q^i|^2\;-\;2\,|H_1|^2\,|Q|^2 \;+ \right. \nonumber \\
 & & 4\,|{H_1^*}^i\,L^i|^2\;-\;2\,|H_1|^2\,|L|^2 \;+\; 4\,|{H_2^*}^i\,Q^i|^2\;
-\;2\,|H_2|^2\,|Q|^2 \;+ \nonumber \\
 & & 4\,|{H_2^*}^i\,L^i|^2\;-\;2\,|H_2|^2\,|L|^2 \;+\; 4\,|{L^*}^i\,Q^i|^2\;
-\;2\,|L|^2\,|Q|^2 \;+ \nonumber \\
 & & \left. |H_1|^4 \;+\;|H_2|^4\;+\;|Q|^4\;+\;|L|^4 \right] \;\;\; , \\
V_S & = & m_{H_1}^2\,|H_1|^2\;+\;m_{H_2}^2\,|H_2|^2\;+\;m_Q^2\,(|t_L|^2\;+\;
|b_L|^2)\;+\;m_t^2\,|t_R|^2\;+ \nonumber \\
 & & m_b^2\,|b_R|^2\;+\;m_L^2\,(|\tau_L|^2\;+\; |\nu_L|^2)\;+\;m_\tau^2\,
|\tau_R|^2 \; + \nonumber \\
& & \left[ \lambda_t\,A_t\,(H_2^+ \,b_L\,\;-\;H_2^0\,t_L)\,t_R \;+\;\lambda_b \,
A_b\, (H_1^0\,b_L \;-\;H_1^-\,t_L)\,b_R\; + \right. \nonumber \\
 & & \left. \lambda_\tau\,A_\tau\, (H_1^0\,\tau_L \;-\;H_1^-\,\nu_L)\, \tau_R \;
 + \; B\,\mu \,(H_2^+\,H_1^- \;-\;H_2^0\,H_1^0) \;\; + \; {\rm{h.c.}} \right] 
\; + \nonumber \\
& & \frac{1}{2}\;\sum_{a=1}^{3}\; M_a \; \bar{\lambda}_a\,\lambda_a \;\;\; .
\label{eq:vs}
\end{eqnarray}
Other than the gauginos $\lambda_a$, all fields in the soft potential are 
scalars. The contributions of the first and second generation sparticles are not
zero - for simplicity we do not write them, but they are identical in form to 
those of their third generation counterparts, minus all terms proportional to 
Yukawa couplings, which are set to zero. The one-loop contributions, at a 
renormalisation scale $M$, are given by 
\begin{equation}
\Delta V_1 \, =\, \sum_\alpha \, \frac{n_\alpha}{64\pi^2}\,M_\alpha^4\, \left(\,
\log \frac{M_\alpha^2}{M^2}\, - \, \frac{3}{2}\,\right) \;\;\; ,
\label{eq:v1cor}
\end{equation}
where the $M_\alpha$ are the (tree-level) masses of each particle of spin
$s_\alpha$ and $n_\alpha = (-1)^{2s_\alpha}\,(2s_\alpha +1)\,C_\alpha\,
Q_\alpha$. $C_\alpha$ is the number of colour degrees of freedom (1 for 
colourless particles, 3 for all others except the gluinos, which, being an
$SU(3)_C$ octet, correspond to a factor of 8) and $Q_\alpha$ is 2 for charged
particles, 1 for chargeless ones. Thus the factor $n_\alpha$ accounts for the
multiplicity of each particle. When the neutral components of $H_1$ and $H_2$ 
acquire vevs $v_1/\sqrt{2}$ and $v_2/\sqrt{2}$, we obtain the ``real" minimum, 
and the tree-level potential $V_0$ may be written as
\begin{equation}
V_0 \; = \;\frac{1}{2} \left( m_1^2\,v_1^2\;+\;m_2^2\,v_2^2 \right) \;-\;B\,\mu
\,v_1\,v_2\;+\;\frac{1}{32} ({g^\prime}^2+g_2^2)\,(v_2^2-v_1^2)^2 \;\;\; ,
\end{equation}
with $m_1^2 = m_{H_1}^2+\mu^2$ and $m_2^2 = m_{H_2}^2+\mu^2$, and where the 
vevs' phases were chosen such that the term with the bilinear coupling is 
negative. Minimising with respect to $v_1$ and $v_2$ we obtain, for the value of
the potential at the minimum, 
\begin{equation}
V_{min}^{MSSM} \; = \; -\frac{1}{32}({g^\prime}^2+g_2^2)\,(v_2^2-v_1^2)^2 
\;\;\; .
\label{eq:vr}
\end{equation}
So that a CCB minimum can exist, we need negative terms in the potential other 
than the one containing $B\,\mu$. These arise from trilinear terms in the 
potential and require non-zero vevs for at least two fields other than 
$H_1^0$ and $H_2^0$. As a working example, consider the case where the fields 
$\tau_L$ and $\tau_R$ have vevs ($l/\sqrt{2}$ and $\tau/\sqrt{2}$ respectively).
The vacuum tree-level potential is now given by
\begin{eqnarray}
V_0 & =& \frac{\lambda_\tau^2}{4} [v_1^2\,(l^2+\tau^2) + l^2\,\tau^2] \;-\;
\frac{\lambda_\tau}{\sqrt{2}}\,(A_\tau\,v_1 + \mu\,v_2)\,l\,\tau \;+ \; 
\frac{1}{2} \,( m_1^2\,v_1^2 \; + \; m_2^2\,v_2^2\; + \; m_L^2\,l^2 \; + 
\nonumber \\
 & & m_\tau^2\,\tau^2 )\;-\;B\,\mu\,v_1\, v_2 \;+\; \frac{g^{\prime 2}}{32}\,
(v_2^2-v_1^2
-l^2+2\,\tau^2)^2 \; +\; \frac{g_2^2}{32}\,(v_2^2-v_1^2+l^2)^2\;\;\; ,
\label{eq:vc}
\end{eqnarray} 
where we chose the phases of the vevs so that the terms with bilinear and 
trilinear couplings are negative~\footnote{Depending on the signs of $A_\tau$,
$\mu$ and $B$, this isn't always possible. This is however a detail, and it 
doesn't have any impact on the conclusions we will draw.}. With four different
vevs in the potential, an analytical study of CCB is very difficult if not 
impossible. Following~\cite{gu} we express all vevs in terms of, say, $v_1$, so
that $l = \alpha\,v_1$, $v_2 = \beta\,v_1$ and $\tau = \gamma\,v_1$, with 
constant and positive parameters $\alpha$, $\beta$, $\gamma$. The 
potential~\ref{eq:vc} becomes a quartic polynomial in $v_1$ and its minimisation
a trivial exercise. Comparing the value of the CCB minimum with that of 
eq.~\ref{eq:vr} and requiring this to be smaller, we find bounds on the MSSM
parameters as function of $\alpha, \ldots$~\cite{cas}. The vevs are found to be
of order $v \sim |A_\tau|/\lambda_\tau$, and the potential $V \sim -|A_\tau|^4/
\lambda_\tau^2$. An important observation throughout is the choice of the 
renormalisation scale at which this calculation is made: as first shown 
in~\cite{gam}, the tree-level vevs are only reliable at a renormalisation scale
of the order of the largest mass present in the one-loop 
contributions~\ref{eq:v1cor}. For our CCB potential, we expect this to be of 
order $M \sim \rm{max}(\lambda_t\, ,\, g_2) |A_\tau|/4 \lambda_\tau$~\cite{cas}.
However, the highest mass present in the MSSM potential will be of the order of 
the gaugino mass, $M_G$, in principle different from $M$. This line of thought 
therefore dictates the two potentials should be compared at different 
renormalisation scales. Now, the work of Gamberini {\em et al} refers to the 
MSSM and demonstrates quite clearly that the tree-level vevs can be trusted in a
given range of the renormalisation scale. Meaning, that in that range the 
one-loop contributions to the minimisation conditions, to wit
\begin{equation}
\sum_\alpha \, \frac{n_\alpha}{32\pi^2}\,M_\alpha^2\,\frac{\partial M_\alpha^2}{
\partial v_i}\, \left(\, \log \frac{M_\alpha^2}{M^2}\, - \, 1\,\right) \;\;\; ,
\label{eq:cont}
\end{equation}
are not significant. It does not, however, say anything about the size of the 
one-loop contributions to the potential themselves. We expect them to be small, 
given that the choice of $M$ limits the size of the logarithmic terms, but 
notice this is not guaranteed. Another and stronger argument for the inclusion 
of the one-loop potential in CCB calculations comes from ref.~\cite{ford}. Their
results are thus summarised: the sum $V_0 + \Delta V_1$ is {\em not} one-loop
renormalisation group invariant. Instead, the full RGE invariant effective 
potential is given by
\begin{equation}
V(M,\lambda_i,\phi_j) \; =\; \Omega(M,\lambda_i) \;+\; V_0(\lambda_i,\phi_j) \;
+\; \Delta V_1 (M,\lambda_i,\phi_j)\; +\; O(h^2) \;\;\; ,
\label{eq:om}
\end{equation}
where $\lambda_i$ stands for all couplings and masses of the theory and $\phi_j$
for all its fields. The function $\Omega$ depends in principle on $M$ - 
implicitly or explicitly - and $\lambda_i$, and receives contributions from all 
orders of perturbation theory. It is, however, field-independent~\footnote{Which
of course means it usually has no effect whatsoever in particle phenomenology, 
given all quantities of physical interest are given by derivatives of $V$ with 
respect to the fields.}. The authors of~\cite{ford} give examples of explicit 
calculations of $\Omega$ for different models and its importance. Notice now 
that the only difference between the ``real" MSSM potential and the CCB one is 
that some fields have vevs, others do not - in other words, the two potentials 
correspond to two different sets of values for the fields in the theory. But the
function $\Omega$ is field-independent and is thus the same, CCB or not. So, 
when we compare $V^{MSSM}$ and $V^{CCB}$ at different renormalisation scales, we
are not taking into account the contribution arising from $\Omega$. We are led 
to the conclusion that, to eliminate $\Omega$ altogether from the calculation, 
we must compare $V^{MSSM}$ and $V^{CCB}$ at the same renormalisation scale. 
Also, notice that eq.~\ref{eq:om} ensures that if $V^{MSSM} > V^{CCB}$ for a 
given scale $M$, the inequality holds for all scales. Now, the rationale we gave
for comparing both potentials at different scales was the possibility of 
neglecting the complex one-loop contributions. The existence of $\Omega$ forcing
a comparison at equal $M$, we conclude that, for one of the potentials at least,
the full one-loop vevs must be used. By consistency we should then work with the
value of the potential correct to one-loop as well. Unfortunately, this way we 
abandon all hope of obtaining analytical CCB bounds, given the complexity of the
one-loop contributions. We now present their calculation in the CCB case. 
Non-zero vevs will be considered only for the real components of the fields 
$H_1^0$ ($v_1/\sqrt{2}$), $H_2^0$ ($v_2/\sqrt{2}$), $\tau_L$ ($l/ \sqrt{2}$) 
and $\tau_R$ ($\tau/\sqrt{2}$). This choice should be favourable to CCB in that 
$m_L$ and $m_\tau$ are generally the smallest of the soft masses and, because no
squarks have vevs, there are no $SU(3)$ D-terms, thus limiting the positive 
contributions to the potential. 
Also, $\lambda_\tau$ being typically quite small, a CCB minimum, should it 
exist, ought to be quite deep. The simplest sparticle mass is that of the 
gluinos ($n_{\tilde{g}} = -16$), it is simply $M_3$, from eq.~\ref{eq:vs}. The 
gluino contribution to $V_1$ is thus field-independent and could be included in 
the $\Omega$ function in eq.~\ref{eq:om} - but in any case where colour breaking
does occur the gluinos would mix with other fields, so, we prefer to single out 
this contribution~\footnote{Of course, given that $V^{CCB}$ and $V^{MSSM}$ will 
be compared at the same renormalisation scale, the gluino contribution will be 
identical in both potentials and have no impact on CCB.}. Also very simple are 
the contributions from the top and bottom quarks ($n_t = n_b = -12$), with their
squared masses given, as usual, by $\lambda_t^2\,v_2^2/2$ and $\lambda_b^2\,
v_1^2/2$ respectively. For the stop ($n_1 = n_2 = 6$), we have
\begin{eqnarray}
[M^2_{\tilde{t}}] & = & \left[ \begin{array}{cc} t_{11} & t_{12} \\
t_{21} & t_{22} \\ \end{array} \right] \;\;\; 
\end{eqnarray}
with
\begin{eqnarray}
t_{11} \; =\; \frac{\partial^2 V}{\partial t_L\partial t_L^*} & = & m_Q^2 \;+\;
\frac{1}{2}\,\lambda_t^2\,v_2^2 \;+\;\frac{{g^\prime}^2}{24}\,(v_2^2-v_1^2-l^2
+2\,\tau^2)\;-\;\frac{g_2^2}{8}\,(v_2^2-v_1^2+l^2) \nonumber \\
t_{12} \; =\; \frac{\partial^2 V}{\partial t_L\partial t_R^*} & = & -\,
\frac{\lambda_t}{\sqrt{2}}\,(A_t\,v_2\;-\;\mu\,v_1) \;\; = \;\; t_{21} 
\nonumber \\            
t_{22} \; =\; \frac{\partial^2 V}{\partial t_R\partial t_R^*} & = & m_t^2 \;+\;
\frac{1}{2}\,\lambda_t^2\,v_2^2 \;-\;\frac{{g^\prime}^2}{6}\,(v_2^2-v_1^2-l^2
+2\,\tau^2) \;\;\; .
\end{eqnarray}
The equality $t_{12}=t_{21}$ arises from the fact we are considering all 
parameters in the potential to be real (the same happens for all the other mass 
matrices, who are thus symmetric). For the sbottom ($n_1 = n_2 = 6$), 
\begin{eqnarray}
[M^2_{\tilde{b}}] & = & \left[ \begin{array}{cc} b_{11} & b_{12} \\
b_{12} & b_{22} \\ \end{array} \right] \;\;\;
\end{eqnarray}
with
\begin{eqnarray}
b_{11} \; =\; \frac{\partial^2 V}{\partial b_L\partial b_L^*} & = & m_Q^2 \;+\;
\frac{1}{2}\,\lambda_b^2\,v_1^2 \;+\;\frac{{g^\prime}^2}{24}\,(v_2^2-v_1^2-l^2
+2\,\tau^2)\;+\;\frac{g_2^2}{8}\,(v_2^2-v_1^2+l^2) \nonumber \\
b_{12} \; =\; \frac{\partial^2 V}{\partial b_L\partial b_R^*} & = & -\,
\frac{\lambda_b}{\sqrt{2}}\,(A_b\,v_1\;-\;\mu\,v_2) \nonumber \\
b_{22} \; =\; \frac{\partial^2 V}{\partial b_R\partial b_R^*} & = & m_b^2 \;+\;
\frac{1}{2}\,\lambda_b^2\,v_1^2 \;+\;\frac{{g^\prime}^2}{12}\,(v_2^2-v_1^2
-l^2+2\,\tau^2) \;\;\; .
\end{eqnarray}
The masses for the first and second generations squarks are obtained from 
these matrices by setting the Yukawas to zero and replacing the soft masses by
the appropriate ones. The squarks are degenerate for these two generations 
(assuming universality of the soft masses - but these results can be trivially
generalised for non-universal models) and, for the up-type squarks ($n_1 = 
n_2 = 2 \times 6$ - the extra factor of two accounts for both generations), we 
have
\begin{eqnarray}
M^2_{\tilde{u}_1} & = & m_R^2 \;+\; \frac{{g^\prime}^2}{24}\,(v_2^2-v_1^2-l^2
+2\,\tau^2)\;-\;\frac{g_2^2}{8}\,(v_2^2-v_1^2+l^2) \nonumber \\
M^2_{\tilde{u}_2} & = & m_u^2 \;-\;\frac{{g^\prime}^2}{6}\,(v_2^2-v_1^2-l^2+
2\,\tau^2) \;\;\; .
\end{eqnarray}
For the down-type squarks ($n_1 = n_2 = 2 \times 6$),
\begin{eqnarray}
M^2_{\tilde{d}_1} & = & m_R^2 \;+\;\frac{{g^\prime}^2}{24}\,(v_2^2-v_1^2-l^2
+2\,\tau^2)\;+\;\frac{g_2^2}{8}\,(v_2^2-v_1^2+l^2) \nonumber \\
M^2_{\tilde{d}_2} & = & m_d^2 \;+\;\frac{{g^\prime}^2}{12}\,(v_2^2-v_1^2-l^2
+2\,\tau^2) \;\;\; .
\end{eqnarray}
Likewise the first and second generation sleptons ($n_1 = n_2 = 2 \times 2$) 
will have degenerate masses given by
\begin{eqnarray}
M^2_{\tilde{e}_1} & = & m_N^2 \;-\;\frac{{g^\prime}^2}{8}\,(v_2^2-v_1^2-l^2
+2\,\tau^2)\;+\;\frac{g_2^2}{8}\,(v_2^2-v_1^2+l^2) \nonumber \\
M^2_{\tilde{e}_2} & = & m_e^2 \;+\;\frac{{g^\prime}^2}{4}\,(v_2^2-v_1^2-l^2+
2\,\tau^2) \;\;\; .
\end{eqnarray}
In these expressions, $m_R$, $m_u$, $m_d$, $m_N$ and $m_e$ are the 
lower-generation equivalents of $m_Q$, $m_t$, $m_b$, $m_L$ and $m_\tau$ 
respectively. Finally, the first and second generation sneutrinos ($n = 2 \times
2$), also degenerate, have a mass given by
\begin{eqnarray}
M^2_{\tilde{\nu}_e} & = & m_N^2 \;-\;\frac{{g^\prime}^2}{8}\,(v_2^2-v_1^2-l^2
+2\,\tau^2)\;-\;\frac{g_2^2}{8}\,(v_2^2-v_1^2+l^2) \;\;\; .
\end{eqnarray}
The masses of the gauge bosons are easily determined (see, for example, 
\cite{bai}). For the charged gauge bosons (we still call them W's; $n = 6$) we 
have
\begin{equation}
M^2_W \;\; = \; \; \frac{1}{4}\;g_2^2\;(v_1^2+v_2^2+l^2) \;\;\;.
\end{equation}
The mass matrix of the neutral gauge bosons ($n_1 = n_2 = 3$) is given by
\begin{eqnarray}
[M^2_{G^0}] & = & \left[ \begin{array}{cc} G_{11} & G_{12} \\
G_{12} & G_{22} \\ \end{array} \right] \;\;\;
\end{eqnarray}
with
\begin{eqnarray}
G_{11} & = & g_2^2\,\sin^2\,\theta_W\,l^2\;+\;{g^\prime}^2\,\tau^2
\nonumber \\
G_{12} & = & \frac{g_2^2}{2}\,\tan\,\theta_W\,\cos (2\theta_W)\,l^2
\nonumber \\
G_{22} & = & \frac{g_2^2}{4\cos^2\theta_W} \;\left[ v_1^2+v_2^2+
\cos^2(2\theta_W)\,l^2 \right] \;\;\; .
\end{eqnarray}
Notice that if $l = \tau = 0$ this matrix produces the usual photon and 
$Z$ masses. The remaining sparticles' mass matrices are complicated by mixing
between fields. Even when there is no CCB some mixing occurs - the charginos
are after all the result of the mixing between the charged gauginos and  
fermionic partners of the Higgses, the neutralinos a similar mix but between
neutral fields - but with the vevs $l$ and $\tau$ in the game, mixing between  
neutral and charged particles occurs. In the case of the charginos, for 
instance, the interaction term in the Lagrangian between the gauge
and matter multiplets, plus the Yukawa interactions in the superpotential, 
cause a mixing with the $\tau$ neutrino (see, for example,
\cite{ros}), $\bar{\nu}_L$ (we use the bar so as to not confuse this field with 
$\nu_L$, one of the components of the scalar doublet $L$) so that, if 
$\lambda^\pm$ are the charged gauginos and $\tilde{H}^\pm$ the fermionic
partners of the charged Higgs fields, the chargino mass matrix will be given by
(the columns, from left to right, correspond to the fields $\lambda^+$, 
$\tilde{H}^+_2$,  $\lambda^-$, $\tilde{H}^-_1$ and $\bar{\nu}_L$)
\begin{equation}
[M_{\chi^\pm}] \;\; =\;\; \frac{1}{2}\;
\left(\begin{array}{ccccc} 0 & 0 & M_2 & -\frac{1}{\sqrt{2}} \,g_2\,v_1 & 0 \\
0 & 0 & \frac{1}{\sqrt{2}} \,g_2\,v_2 & -\mu & 0 \\ 
M_2 & \frac{1}{\sqrt{2}} \,g_2\,v_2 & 0 & 0 & \frac{1}{\sqrt{2}} \,g_2\,l \\ 
-\frac{1}{\sqrt{2}} \,g_2\,v_1 & -\mu & 0 & 0 & - \frac{1}{\sqrt{2}} \,
\lambda_\tau\,\tau \\ 0 & 0 & \frac{1}{\sqrt{2}} \,g_2\,l & - \frac{1}{\sqrt{2}}
 \,\lambda_\tau \,\tau & 0 \\ \end{array} \right) \;\;\; .
\end{equation}
This matrix has one zero eigenvalue (corresponding to the neutrino which, 
despite the mixing, does not gain a mass in this scenario) and the squared 
masses of the two charginos ($n_1 = n_2 = -4$) are determined by a simple 
quadratic equation. The neutralinos ($n_{1 \ldots 6} = -2$) now include mixing 
with the tau fermion, and their mass matrix is (from left to right, the entries 
of the matrix are $\tilde{B}$, $\tilde{W}^3$, $\tilde{H}^0_1$, $\tilde{H}^0_2$, 
$\bar{\tau}_L$ and $\bar{\tau}_R$)
\begin{equation}
[M_{\chi^0}] \;\; =\;\; \frac{1}{2}\; \left(\begin{array}{cccccc} M_1 & 0 & 
-\,\frac{1}{2}\,g^\prime \,v_1 & \frac{1}{2}\,g^\prime \,v_2 & -\,
\frac{1}{2}\,g^\prime \, l & g^\prime \,\tau \\
0 & M_2 & \frac{1}{2}\,g_2 \,v_1 & -\frac{1}{2}\,g_2 \,v_2 & \frac{1}{2}
\,g_2 \,l & 0 \\
-\,\frac{1}{2}\,g^\prime \,v_1 & \frac{1}{2}\,g_2 \,v_1 & 0 & -\mu & 
\frac{1}{\sqrt{2}}\,\lambda_\tau\, \tau & \frac{1}{\sqrt{2}}\,\lambda_\tau\, l 
\\
 \frac{1}{2}\,g^\prime \,v_2 &  -\frac{1}{2}\,g_2 \,v_2 & -\mu & 0 & 0 & 0 \\
-\,\frac{1}{2}\,g^\prime\,l & \frac{1}{2}\,g_2\,l & \frac{1}{\sqrt{2}}\,
\lambda_\tau\, \tau & 0 & 0 & \frac{1}{\sqrt{2}}\,\lambda_\tau\, v_1 \\ 
 g^\prime \,\tau & 0 & \frac{1}{\sqrt{2}}\,\lambda_\tau\, l & 0 & 
\frac{1}{\sqrt{2}}\,\lambda_\tau\, v_1 & 0 \\ \end{array} \right) \;\;\; .  
\end{equation}
The charged Higgs ($n_{1 \ldots 3} = 2$) fields of the theory mix with the 
$\tau$ sneutrinos, so that their combined mass matrix is (from left to right, 
$H_1^-$, $H_2^+$ and $\nu_L$)
\begin{equation}
[M^2_{H^\pm \nu_\tau}] \;\; =\;\; \left(\begin{array}{lll} a_\pm & 
b_\pm & d_\pm \\ b_\pm & c_\pm & e_\pm \\ d_\pm &
e_\pm & f_\pm \\ \end{array}\right) \;\;\;,
\end{equation}
with
\begin{eqnarray}
a_\pm & = & 
m_1^2\;+\; \frac{\lambda_\tau^2}{2}\,\tau^2\;-\;\frac{{g^\prime}^2}{8}\,(v_2^2
-v_1^2-l^2+2\,\tau^2)\;+\;\frac{g_2^2}{8}\,(v_2^2+v_1^2+l^2) \nonumber \\
b_\pm & = &
B\,\mu\; + \; \frac{g_2^2}{4}\,v_1\,v_2 \nonumber \\
c_\pm & = &
m_2^2 \;+\; \frac{{g^\prime}^2}{8}\,(v_2^2-v_1^2-l^2+2\,\tau^2)
\;+\;\frac{g_2^2}{8}\,(v_2^2+v_1^2-l^2) \nonumber \\
d_\pm & = & 
-\frac{\lambda_\tau}{\sqrt{2}}\,A_\tau\,\tau \;-\;\left(\frac{\lambda_\tau^2}{2}
\;-\;\frac{g_2^2}{4}\right)\,v_1 \,l \nonumber \\
e_\pm & = &
-\frac{\lambda_\tau}{\sqrt{2}}\,\mu\,\tau \;+\;\frac{g_2^2}{4}\,v_2 \, l 
\nonumber \\ 
f_\pm & = & m_L^2 \; +\; \frac{\lambda_\tau^2}{2}\,\tau^2\;-\;
\frac{{g^\prime}^2}{8}\,(v_2^2-v_1^2-l^2+2\,\tau^2) \;-\;
\frac{g_2^2}{8}\,(v_2^2-v_1^2- l^2) \;\;\; .
\end{eqnarray}
It is easy to check, using the tree-level minimisation conditions, that this 
matrix has a zero eigenvalue, corresponding to the two Goldstone bosons that 
give mass to the charged $W$ gauge bosons. The pseudoscalars ($n_{1 \ldots 4} = 
1$) in this theory are a mix between the imaginary parts of the neutral 
components of $H_1$, $H_2$ and $\tau_L$ and $\tau_R$. The mass matrix is given 
by 
\begin{equation}
[M^2_{\bar{H^0 \tau}}] \;\; =\;\; \left(\begin{array}{llll} 
a_{\bar{H}} & b_{\bar{H}} & d_{\bar{H}} & e_{\bar{H}} \\ 
b_{\bar{H}} & c_{\bar{H}} & f_{\bar{H}} & g_{\bar{H}} \\ 
d_{\bar{H}} & f_{\bar{H}} & h_{\bar{H}} & i_{\bar{H}} \\ 
e_{\bar{H}} & g_{\bar{H}} & i_{\bar{H}} & j_{\bar{H}} \\
\end{array}\right) \;\;\;,
\end{equation}
where each of the entries of this matrix is a derivative of the form 
$\partial^2 V/\partial Im\,(\phi_i)\,\partial Im\,(\phi_j)$, the fields $\phi_i$
corresponding to, from left to right, $H_1^0$, $H_2^0$, $\tau_L$ and $\tau_R$. 
We have
\begin{eqnarray}
a_{\bar{H}} & = & m_1^2\;+\; 
\frac{\lambda_\tau^2}{2}\,(l^2+\tau^2) \;-\;\frac{{g^\prime}^2}{8}\,
(v_2^2-v_1^2 -l^2+2\,\tau^2) \; -\; \frac{g_2^2}{8}\,(v_2^2-v_1^2+ l^2) 
\nonumber \\
b_{\bar{H}} & = & B\,\mu \nonumber \\
c_{\bar{H}} & = & m_2^2\;+\; \frac{{g^\prime}^2}{8}
\,(v_2^2-v_1^2 -l^2+2\,\tau^2)\;+\; \frac{g_2^2}{8}\,(v_2^2-v_1^2+l^2) 
\nonumber \\ 
d_{\bar{H}}  & = & -\,\frac{\lambda_\tau}{\sqrt{2}}
\,A_\tau\, \tau \nonumber \\
e_{\bar{H}} & = & -\,\frac{\lambda_\tau}{\sqrt{2}}
\,A_\tau\, l \nonumber \\
f_{\bar{H}} & = & -\,\frac{\lambda_\tau}{\sqrt{2}}
\,\mu \, \tau \nonumber \\
g_{\bar{H}} & = & -\,\frac{\lambda_\tau}{\sqrt{2}}
\,\mu \, l \nonumber \\
h_{\bar{H}} & = & m_L^2\;+\; \frac{\lambda_\tau^2}{2}\,
(v_1^2+\tau^2)
\;-\;\frac{{g^\prime}^2}{8}\,(v_2^2-v_1^2-l^2+2\,\tau^2)\nonumber \\
 & & +\; \frac{g_2^2}{8}\,(v_2^2-v_1^2+l^2) \nonumber \\
i_{\bar{H}} & = & \frac{\lambda_\tau}{\sqrt{2}}
\, (\mu\, v_2 \; -\; A_\tau\, v_1) \nonumber \\
j_{\bar{H}} & = & m_\tau^2\;+\; 
\frac{\lambda_\tau^2}{2}\,(v_1^2+l^2)
\;+\;\frac{{g^\prime}^2}{4}\,(v_2^2-v_1^2-l^2+2\,\tau^2) \;\;\; .
\end{eqnarray}
Using the tree-level minimisation conditions, we see that this matrix has 
two zero eigenvalues, corresponding to the Goldstone bosons that give mass to
the neutral gauge bosons - we thus have four Goldstone bosons, one charged and 
two neutral, exactly as we should have, given we are fully breaking the 
$SU(2)_W \times U(1)_Y$ gauge symmetry. The ``Higgs scalars" ($n_{1 \ldots 4} = 
1$) of this theory are a mix of the real parts of the neutral components of 
$H_1$ and $H_2$ and $\tau_L$ and $\tau_R$. The mass matrix is 
\begin{equation}
[M^2_{H^0 \tau}] \;\; =\;\; \left(\begin{array}{llll}
a_H & b_H & d_H & e_H \\
b_H & c_H & f_H & g_H \\
d_H & f_H & h_H & i_H \\
e_H & g_H & i_H & j_H \\
\end{array}\right) \;\;\;,
\end{equation}
where each entry is now given by $\partial^2 V/\partial Re\,(\phi_i)\,\partial 
Re\,(\phi_j)$ and we have
\begin{eqnarray}
a_H & = & m_1^2\;+\;
\frac{\lambda_\tau^2}{2}\,(l^2+\tau^2) \;-\;\frac{{g^\prime}^2}{8}\,
(v_2^2-3\,v_1^2 -l^2+2\,\tau^2)\;-\; \frac{g_2^2}{8}\,(v_2^2-3\,v_1^2+l^2) 
\nonumber \\
b_H & = & -\,B\,\mu \;-\; \frac{1}{4}\,(
{g^\prime}^2+g_2^2)\,v_1\,v_2 \nonumber \\
c_H & = & m_2^2\;+\; \frac{{g^\prime}^2}{8}
\,(3\,v_2^2-v_1^2-l^2+2\,\tau^2)\;+\; \frac{g_2^2}{8}\,(3\,v_2^2-v_1^2+l^2)
\nonumber \\
d_H & = & \frac{\lambda_\tau}{\sqrt{2}}
\,A_\tau\, \tau \;+\; \left[\lambda_\tau^2 \;+\; \frac{1}{4}\,({g^\prime}^2
-g_2^2) \right]\,v_1\,l \nonumber \\
e_H & = & \,\frac{\lambda_\tau}{\sqrt{2}}
\,A_\tau\, l \;+\; \left(\lambda_\tau^2 \;-\; \frac{1}{2}\,{g^\prime}^2
\right)\,v_1\,\tau \nonumber \\
f_H & = & -\,\frac{\lambda_\tau}{\sqrt{2}}
\,\mu \, \tau \;-\; \frac{1}{4}\,({g^\prime}^2
-g_2^2) \,v_2\,l \nonumber \\
g_H & = & -\,\frac{\lambda_\tau}{\sqrt{2}}
\,\mu \, l \;+\;  \frac{1}{2}\,{g^\prime}^2 \,v_2\,\tau \nonumber \\
h_H & = & m_L^2\;+\; \frac{\lambda_\tau^2}{2}\,
(v_1^2+\tau^2) \;-\;\frac{{g^\prime}^2}{8}\,(v_2^2-v_1^2-3\,l^2+2\,\tau^2)
\;+\; \frac{g_2^2}{8}\,(v_2^2-v_1^2+3\, l^2) \nonumber \\
i_H & = & \frac{\lambda_\tau}{\sqrt{2}}
\, (\mu\, v_2 \; -\; A_\tau\, v_1) \;+\; \left(\lambda_\tau^2\;-\;
\frac{{g^\prime}^2}{2}\right) \,l\,\tau \nonumber \\
j_H & = & m_\tau^2\;+\;
\frac{\lambda_\tau^2}{2}\,(v_1^2+l^2) \;+\;\frac{{g^\prime}^2}{4}\,(v_2^2-v_1^2 
-l^2+6\,\tau^2) \;\;\; .
\end{eqnarray}

A simple check we can perform on the validity of these formulas is to calculate 
$Str\,M^2 \,=\, \sum_\alpha \, n_\alpha\,M_\alpha^2$ - because supersymmetry 
is broken in a ``soft" way, this quantity should be field-independent. It is 
not difficult to verify that the mass matrices here presented satisfy this 
condition. Let us now apply these results to a simple example, the MSSM with 
universality (common values for the gaugino and scalar masses and the $A$ 
parameters at the gauge unification scale $M_U \sim 10^{16}$ GeV). Besides their
simplicity there are several reasons these models are of interest: the absence 
of flavour changing neutral currents at the weak scale is a strong indicator of 
universal soft parameters. There is also the evidence that a vast range of 
theories have an universal form at $M_U$ as an infrared fixed point~\cite{ifp}. 
Finally, these models are of phenomenological interest in that their predictions
for the sparticle masses have not been excluded by the current experimental data
(see, for instance, ref.~\cite{euro}). A brief description of our procedure: we 
input at scale $M_Z$ the masses of the gauge bosons, third generation fermions, 
the values of the gauge couplings and $\tan \beta$ (ranging from 2.5 to 10.5). 
Using the two-loop $\beta$-functions, we then find the gauge 
unification scale (unification of $\alpha_1$ and $\alpha_2$ only) at which point
we input the universal soft parameters ($20 \leq M_G \leq 100$ GeV, $10 \leq m_G
\leq 160$ GeV, $-600 \leq A_G \leq 600$ GeV). We then go down to scale $M_S = 
{\rm max}(M_Z, M_G, m_G)$ (of the order of the largest mass in the theory) and 
determine $B$ and $\mu$ from the minimisation of the full one-loop potential 
(see, for instance, \cite{bbo}). At this stage we have about 3200 ``points" of 
the parameter space of the MSSM. At the scale $M = 0.6 \,g_2 |A_\tau|/
\lambda_\tau$ we then solve the CCB minimisation conditions - the set of 
equations $\{ \partial V_0/\partial v_1 = 0 \;,\; \partial V_0/\partial v_2 = 0 
\;,\;  \partial V_0/ \partial l = 0 \;,\; \partial V_0/\partial \tau = 0\}$, 
with $V_0$ from eq.~\ref{eq:vc} - which is equivalent to determining the vevs 
$\{v_1\,,\, v_2\,, \, l\,,\, \tau\}$. Notice that we solve the {\em tree-level} 
minimisation conditions for the CCB potential - we are relying on the work of 
Gamberini {\em et al}, assuming that at the scale $M$ the 
contributions~\ref{eq:cont} are not significative~\footnote{We chose $M$, after 
several attempts, to be the value for which the ratio $r \,=\, {\rm max}
(M_\alpha^2)/M^2$ is closest to $1$. At this scale the one-loop contributions to
the minimisation of $V^{MSSM}$ are expected to be significative, so, they are 
included.}. We then calculate $(V_0 + \Delta V_1)^{CCB}$ and compare it - at the
same scale $M$ - with $(V_0 + \Delta V_1)^{MSSM}$. For much of the parameter 
space considered solving the CCB minimisation conditions is not possible. For 
$\sim 1290$ ``points", we find extrema for which the value of the potential is 
smaller than $(V_0 + \Delta V_1)^{MSSM}$. We can evaluate the importance the
choice of renormalisation scale has - if we choose, for instance, $M = \,g_2 
|A_\tau|/ 4 \lambda_\tau$, we see CCB occurring for 200 more ``points". Notice
how our choice of renormalisation scale is always a field-independent 
one~\cite{ford}. This dependence on $M$ would, one expects, vanish if we had 
performed a full one-loop minimisation of the CCB potential~\cite{eu}.

\begin{figure}[htb]
\epsfysize=8cm
\centerline{\epsfbox{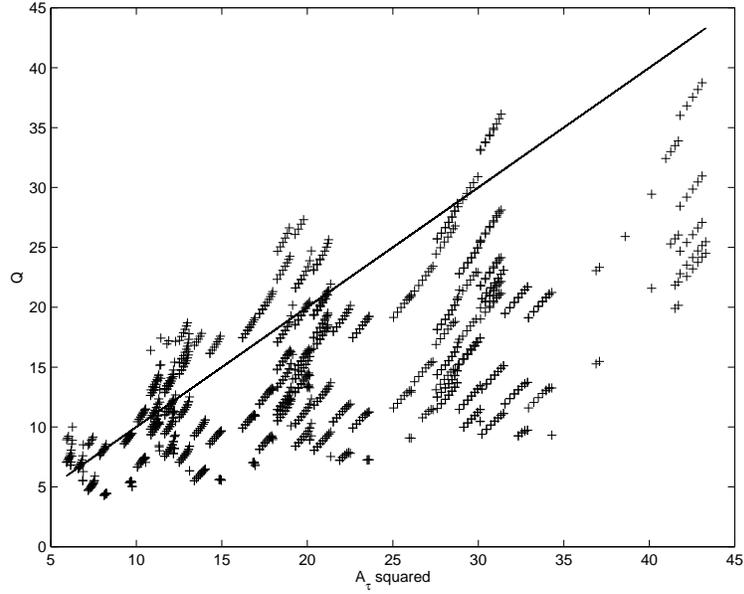}}
\caption{$A_\tau^2$ {\em versus} $Q \,=\, 3\, (m_1^2 \, + \,
m_L^2 \, +\, m_\tau^2)$, for the rejected CCB points at the scale $M$; all 
points below the solid line violate the ``classic" CCB bound of Fr\`ere {\em et
al}.}
\label{fig:cla}
\end{figure}

\begin{figure}[htb]
\epsfysize=8cm
\centerline{\epsfbox{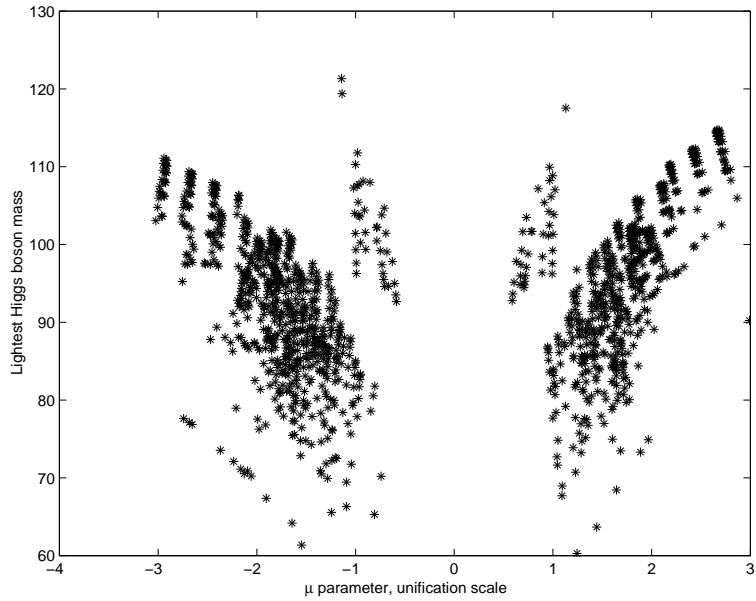}}
\caption{Mass of the lightest Higgs boson {\em versus} the value of the $\mu$   
parameter at the unification scale, for the CCB-rejected points.}
\label{fig:mu}
\end{figure}

These CCB extrema are distributed uniformly through the input values of $\tan 
\beta$, $M_G$ and $m_G$, but concentrate on the high absolute values of $A_G$ -
we only see then occurring for $|A_G| \geq 300$ GeV. This of course reminds us
of the ``classic" CCB bound~\cite{fre}, according to which CCB would occur if
$|A_\tau|^2 \, \geq \,3\, (m_1^2 \, + \, m_L^2 \, +\, m_\tau^2)$. 
Figure~\ref{fig:cla} shows us that the majority of rejected ``points" doesn't 
satisfy this bound. They also 
concentrate in specific values of the $\mu$ parameter, as may be seen in 
fig.~\ref{fig:mu}. Interestingly, the rejected ``points" target higher values of
$M_h$ - notice how few rejections occur for $M_h < 80$ GeV - and do not occur 
for small values of $|\mu|$. This Higgs mass there plotted is that of the 
``real", MSSM, Higgs boson, calculated from the full one-loop mass matrices. 
Another interesting observation is that the values found for the vevs $l$ and 
$\tau$ are very similar. To evaluate the impact of using the full one-loop 
potential and comparing both potentials at the same scale we performed the 
``usual" CCB analysis (i.e, compare the tree-level potential only, at two 
different renormalisation scales) as well. We found CCB happening for $\sim$ 
10\% more ``points" than we have found using the full one-loop potential. 

Some remarks: the CCB extrema we found are {\em not} minima, in that there are
always some quadratic sparticle masses which are negative (specifically, 
$\tilde{b}_2$, the $\tilde{u}$'s and $\tilde{\nu}_e$). It has been 
shown~\cite{rai} that a non-convex effective potential is the result of a bad 
summation of 1PI 
diagrams, and the way to proceed in such situations is to take the {\em convex
envelope} of the potential. A good approximation is to consider the potential
flat where negative second derivatives occur. We point out, however, that the
negative masses found are very small when compared to the typical masses in 
the CCB potential, and can therefore be safely neglected. Also, we must remember
that all that has been said here relates to the possibility of the ``real" 
vacuum being the absolute minimum of the potential. There is however the 
argument~\cite{ccb} that CCB bounds should be determined from requiring the
tunnelling time from the ``real" minimum to the CCB one to be larger than the 
age of the universe. This, of course, leads to a relaxation on the bounds. We do
not undertake this study here, but argue that, in any case, the one-loop 
contributions to the potential would be relevant, in that they determine how
deep the CCB minima/extrema are. The importance of the one-loop contributions
to CCB bounds had already been studied in ref.~\cite{baer}, but comparison with
their results is difficult: they considered only the top-stop contributions to
$\Delta V_1$ (we found that those are very important, but no more so than the 
chargino and ``Higgs scalars" ones) and used a calculational method very 
different from the one here presented.  In conclusion, we calculated a full 
one-loop CCB potential, and argued that the existence of the $\Omega$ function -
that is to say, the RG invariance of the effective potential - forces a 
comparison of the CCB and MSSM potentials at the same renormalisation scale $M$.
Having the actual CCB masses informed our choice of $M$, higher than usually 
considered.  In a simple example, we found a small, but nevertheless sizeable, 
difference between the parameter space rejected in the two CCB approaches here 
delineated. It remains to be seen what that difference might be for other 
models, more complex or with larger parameter space. In light of the (hopefully)
impending discovery of the Higgs boson, a careful reconsideration of CCB bounds 
might be in order. Finally, the computation of the CCB mass matrices opens the 
way to the possibility of performing a full one-loop minimisation of the CCB 
potential. That way one could confirm if their impact in the bounds is indeed 
negligible, and eliminate the renormalisation scale dependence we encountered. 
This work is in progress~\cite{eu}.

\vspace{0.25cm}
{\bf Acknowledgements.} I thank Tim Jones and Giovanni Ridolfi for very useful 
discussions.  This work was supported in part by the European Commission TMR 
program ERB FMRX-CT96-0045.

\vspace{1cm}
\noindent
{\Large {\bf Figure Captions.}}
\begin{description}
\item[Fig.~\ref{fig:cla}.] $A_\tau^2$ {\em versus} $Q \,=\, 3\, (m_1^2 \, + \, 
m_L^2 \, +\, m_\tau^2)$, for the rejected CCB points at the scale $M$; all 
points below the solid line violate the ``classic" CCB bound of Fr\`ere {\em et 
al}.
\item[Fig.~\ref{fig:mu}.] Mass of the lightest Higgs boson {\em versus} the 
value of the $\mu$ parameter at the unification scale, for the CCB-rejected 
points.
\end{description}

\vspace{1cm}

\end{document}